\documentclass[reprint,amsmath,amssymb,aps,superscriptaddress, nofootinbib]{revtex4-2}

\usepackage{graphicx} 

\usepackage{amsmath}
\usepackage[colorlinks=true, allcolors=blue]{hyperref}
\usepackage{cleveref}
\usepackage{bm}
\usepackage{amssymb}
\usepackage{tikz}
\usetikzlibrary{quantikz}
\usepackage{comment}
\usepackage{chemformula}
\usepackage{url}

\usepackage{academicons}
\usepackage{xcolor}

\usepackage{orcidlink}

\usepackage[colorinlistoftodos, textsize=small, color=orange!50]{todonotes}

\begin{document}
\title{Quantum Simulation of Electron Energy Loss Spectroscopy for Battery Materials}

\author{Alexander Kunitsa\orcidlink{0000-0002-3640-8548}} 
\author{Diksha Dhawan\orcidlink{0000-0002-1129-3166}}
\author{Stepan Fomichev\orcidlink{0000-0002-1622-9382}}
\author{Juan Miguel Arrazola\orcidlink{0000-0002-0619-9650}}
\affiliation{Xanadu, Toronto, ON, M5G2C8, Canada}
\author{Minghao Zhang\orcidlink{0009-0002-8303-3942}}
\affiliation{Pritzker School of Molecular Engineering, The University of Chicago, Chicago, USA}
\author{Torin F. Stetina\orcidlink{0000-0002-6279-8455}} 
\affiliation{Xanadu, Toronto, ON, M5G2C8, Canada}

\begin{abstract}
The dynamic structure factor (DSF) is a central quantity for interpreting a vast array of inelastic scattering experiments in chemistry and materials science, but its accurate simulation is a considerable challenge for classical computational methods. In this work, we present a quantum algorithm and an end-to-end simulation framework to compute the DSF, providing a general approach for simulating momentum-resolved spectroscopies. We apply this approach to the simulation of electron energy loss spectroscopy (EELS) in the core-level electronic excitation regime, a spectroscopic technique offering sub-nanometer spatial resolution and capable of resolving element-specific information, crucial for analyzing battery materials. We derive a quantum algorithm for computing the DSF for EELS by evaluating the off-diagonal terms of the time-domain Green's function, enabling the simulation of momentum-resolved spectroscopies. To showcase the algorithm, we study the oxygen K-edge EELS spectrum of lithium manganese oxide (\ch{Li2MnO3}), a prototypical cathode material for investigating the mechanisms of oxygen redox in battery materials. For a representative model of an oxygen-centered cluster of \ch{Li2MnO3} with an active space of 18 active orbitals, the algorithm requires a circuit depth of $3.25\times10^{8}$ T gates, 100 logical qubits, and roughly $10^4$ shots.
\end{abstract}

\maketitle

\section{Introduction}
The development of advanced energy storage systems, such as state-of-the-art lithium-ion batteries~\cite{kim2019lithium}, or more nascent metal-air batteries~\cite{rahman2013high}, is critical for applications spanning portable electronics, electric vehicles, and grid-scale storage. Enhancing battery performance is a multifaceted challenge that requires optimizing the energy density, lifespan, and safety of its core components, such as electrodes and electrolytes. A key theoretical tool in this effort is the dynamic structure factor (DSF), a central quantity in materials science that links theoretical models to experimental spectroscopy data and applies to inelastic scattering experiments most generally~\cite{sturm_dynamic_1993}. The DSF contains all of the information necessary to interpret and quantify a large array of spectroscopic techniques that provide crucial insights into the microscopic dynamics within a material, including inelastic X-ray scattering, inelastic neutron scattering, and electron energy loss spectroscopy (EELS)~\cite{sturm_dynamic_1993,radtke_energy_2011,egerton_electron_2011}. While an accurate simulation of the DSF is important for interpreting the complex molecular structure-property relationships from experimental data, it can pose a considerable challenge for classical simulation methods, especially for strongly correlated materials.~\cite{ikeno_basics_2017,ikeno_multiplet_2009,luder_theory_2017,mizoguchi_theoretical_2010,de_groot_high-resolution_2001,de_groot_2p_2021}

An experimental technique, known as STEM/EELS, combines the sub-nanometer spatial resolution of scanning transmission electron microscopy (STEM) with EELS to probe local features like defects and interfaces. Specifically, analyzing the energy loss near-edge structure (ELNES) provides detailed local chemical state information, such as oxidation states and bonding environments, that are crucial for understanding and improving battery materials.~\cite{yu_stem-eels_2021}. However, the complicated fine structures of ELNES spectra, especially in metal-oxide cathodes, present significant interpretational challenges, necessitating accurate computational methods for spectral fingerprinting. 

Accurate simulation of ELNES can provide reference spectra that are essential for interpreting experimental data, including local molecular structure and redox states. Current state-of-the-art classical methods for simulating ELNES face serious limitations. Approaches using density functional theory (DFT)~\cite{ikeno_basics_2017,tait_simulation_2016,gao_core-level_2009}, time-dependent DFT (TDDFT)~\cite{kasper_ab_2020,Besley2010,lopata_linear-response_2012}, and the Bethe-Salpeter Equation~\cite{vinson_bethe-salpeter_2011,ikeno_basics_2017,yao_all-electron_2022} are popular due to their computational efficiency, but ensuring accurate modeling of the requisite core-hole effects and electronic correlations in the ground and excited states requires more computationally demanding methods. For systems with strong electronic correlation, wavefunction-based restricted active space (RAS) approaches~\cite{pinjari_restricted_2014,kasper_ab_2020,bokarev_theoretical_2020} are more suitable, but their high computational cost restricts them to small active spaces. For instance, highly accurate calculations have been limited to active spaces of 13 or fewer orbitals~\cite{pinjari_cost_2016}. This can lead to inaccuracies in predicted spectra if important orbitals cannot be included in the active space due to simulation cost constraints.
\begin{figure*}
    \centering
    \includegraphics[width=\textwidth]{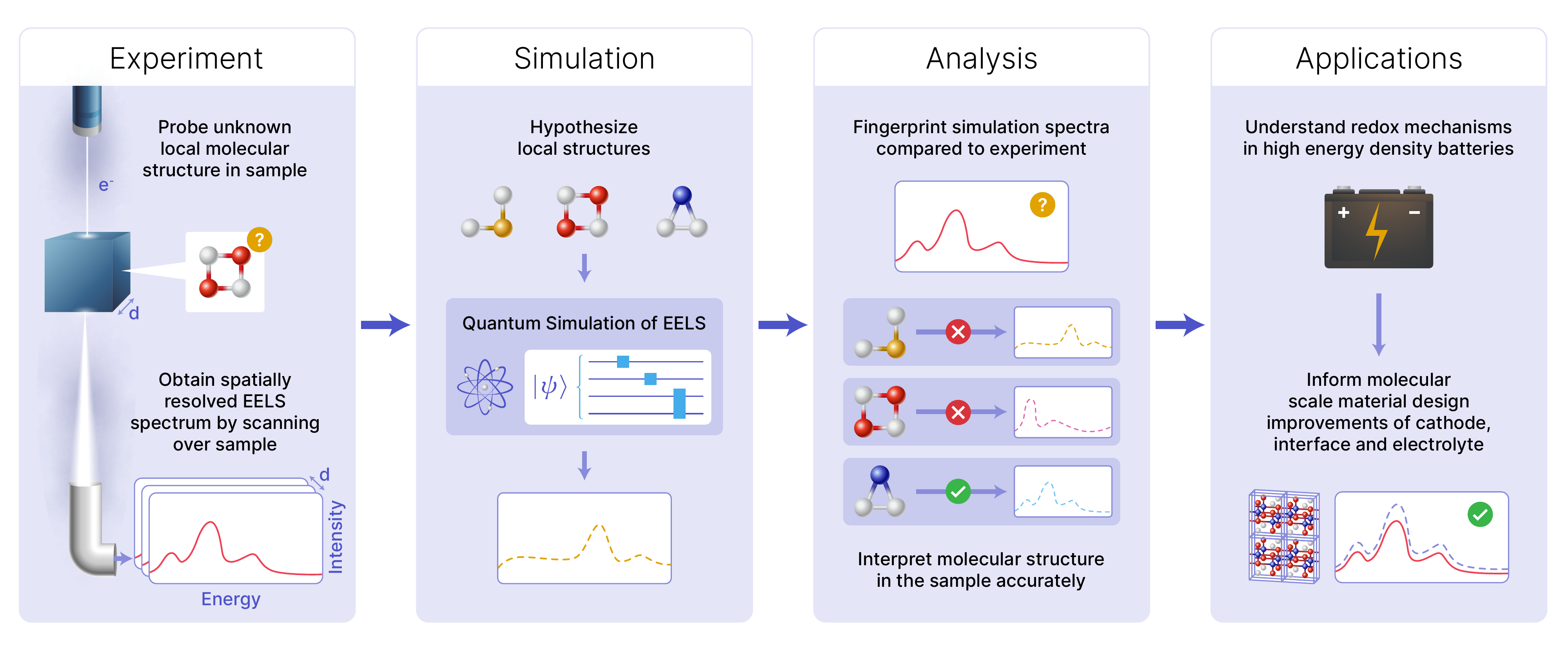}
\caption{A high-level end-to-end application pipeline for the quantum simulation of EELS spectra and structural analysis of local defects in battery cathode materials. First, experimental EELS spectra are obtained by scanning through a section of a (thin) sample with a focused electron beam in a (scanning) transmission electron microscope (STEM) and recording the energy distribution of scattered electrons. An atomically resolved STEM image of the same section is acquired as well. To interpret the spectra, specifically in the case of ELNES, one hypothesizes relevant local structures and computes the ELNES cross section for each, establishing their spectral fingerprints. Subsequently, the experimental spectra are compared against the theory. In complicated cases, where several local structures contribute to the spectra, advanced deconvolution techniques are used. Such analysis allows one to infer properties such as spatial distributions of oxidation states and local coordination environments, providing insights into the chemistry of heterogeneous materials. In particular, this could be used to reveal the details of redox processes in high-energy density batteries and inform new material designs working toward more stable, efficient, and safe energy storage.} 
    \label{fig:hero}
\end{figure*}

In this work, we develop a quantum simulation framework for the DSF, and subsequently focus on EELS simulation for electronic core excited states in the high energy (100s to 1000s of eV) regime ~\cite{radtke_energy_2011,egerton_electron_2011,ikeno_basics_2017}. As mentioned previously, this energy regime is sometimes more specifically referred to as ELNES, but since our approach works for EELS more generally, we will use the terms EELS and ELNES interchangeably unless otherwise noted. The primary contribution of our work is a quantum algorithm for computing the DSF. We envision the algorithm as an integral component of a corresponding workflow to interpret experimental EELS data from battery materials, schematically depicted in Fig.~\ref{fig:hero}. We demonstrate that EELS spectra can be simulated on a quantum computer at a cost comparable to that of X-ray absorption spectroscopy (XAS) calculations, as done in the recent work of Ref.~\cite{fomichev_fast_2025}. As the main technical result, we introduce a method that is a significant generalization of previous time-domain quantum algorithms for computing linear absorption spectra of molecules and materials. Specifically, we present a novel algorithm that directly computes the off-diagonal terms of the time-domain Green's function (the dipole-dipole correlation function between two different Cartesian components). This is accomplished by designing a quantum circuit that prepares and propagates a superposition of two distinct dipole-rotated initial states with any pair of Cartesian components, which in turn allows for the measurement of the required expectation values. This contribution is the central component that makes the quantum simulation of orientation-dependent spectroscopies feasible,  generalizing prior work and providing a pathway for simulating a broader class of momentum-resolved spectroscopies and inelastic scattering experiments. 

To demonstrate our quantum algorithm's utility, we apply it to one of the key challenges in materials science: the computational modeling of EELS for advanced battery materials. We showcase our method by simulating and providing quantum resource estimates for the oxygen K-edge ELNES spectrum of lithium manganese oxide (\ch{Li2MnO3}), a prototypical cathode material where understanding complex oxygen redox mechanisms is critical for designing next-generation batteries.

To ensure our simulations are representative of the local chemical environment, we performed a careful construction of the electronic structure Hamiltonian. We derived a cluster model from the experimental crystal structure that correctly reproduces the essential physics, including the local oxygen environment and the antiferromagnetic ordering of the manganese centers, features which are crucial for a meaningful simulation of spectra. Using this model, we provide preliminary quantum resource estimates on the order of $10^{8} - 10^{9}$ T gates for the largest circuit depth, with roughly $10^{4}$ shots per circuit needed to resolve the spectrum using a number of logical qubits on the order of 100. These resource estimates that are reported in Tab.~\ref{tab:resources}, and discussed in detail in \Cref{sec:app}, were obtained using highly optimized quantum subroutines, including the Sum of Slaters (SOS) state preparation method~\cite{fomichev_initial_2024}, and Trotter product formulas for the Hamiltonian time evolution~\cite{childs_theory_2021,fomichev_fast_2025}. Our results demonstrate the feasibility of accurately computing EELS for classically intractable electronic structure models on fault-tolerant quantum computers.

\begin{table*}[t]
\centering
\begin{tabular}{c c c c c c}
\hline
\multicolumn{2}{c}{} & \multicolumn{2}{c}{Algorithm} & \multicolumn{2}{c}{Largest Circuit} \\
\hline\hline
 $N_a$ & Logical qubits & T gates & Active Volume & T gates & Active Volume \\
\hline
$14$ & $92$ & $1.54 \times 10^{12}$ & $4.13 \times 10^{13}$ & $1.67 \times 10^{8}$ & $4.46 \times 10^{9}$ \\
$16$ & $96$ & $2.29 \times 10^{12}$ & $6.16 \times 10^{13}$ & $2.47 \times 10^{8}$ & $6.65 \times 10^{9}$ \\
$18$ & $100$ & $3.25 \times 10^{12}$ & $8.77 \times 10^{13}$ & $3.51 \times 10^{8}$ & $9.48 \times 10^{9}$ \\
$20$ & $104$ & $4.45 \times 10^{12}$ & $1.20 \times 10^{14}$ & $4.80 \times 10^{8}$ & $1.30 \times 10^{10}$ \\
$22$ & $108$ & $5.91 \times 10^{12}$ & $1.60 \times 10^{14}$ & $6.39 \times 10^{8}$ & $1.73 \times 10^{10}$ \\
$24$ & $112$ & $7.67 \times 10^{12}$ & $2.08 \times 10^{14}$ & $8.29 \times 10^{8}$ & $2.25 \times 10^{10}$ \\
$26$ & $116$ & $9.75 \times 10^{12}$ & $2.65 \times 10^{14}$ & $1.05 \times 10^{9}$ & $2.86 \times 10^{10}$ \\
$28$ & $120$ & $1.22 \times 10^{13}$ & $3.31 \times 10^{14}$ & $1.32 \times 10^{9}$ & $3.58 \times 10^{10}$ \\
$30$ & $124$ & $1.50 \times 10^{13}$ & $4.08 \times 10^{14}$ & $1.62 \times 10^{9}$ & $4.40 \times 10^{10}$ \\
\hline
\end{tabular}
\caption{Resource estimates for different numbers of spatial orbitals ($N_a$) for resolving EELS spectra from quantum simulation based on the \ch{Li2MnO3} molecular cluster model detailed in \Cref{sec:app}. T gate and active volume values under the ``Largest Circuit" heading refer to the deepest Hadamard test circuit required to compute the dynamic structure factor $S(q,\omega)$ for $q = (1,1,1)$. The values under the ``Algorithm" heading account for quantum resources across all Hadamard test circuits using $10^4$ samples. Active volume was computed as described in Ref.~\cite{litinski_active_2022}. Further details are discussed in \Cref{sec:app}.}
\label{tab:resources}
\end{table*}

The remainder of this paper is organized as follows: \Cref{sec:background} introduces the dynamic structure factor as a tool for computing the double differential cross section in momentum-resolved spectroscopy, and reviews the basic theory of EELS, highlighting the time-domain approach which is amenable to implementation on quantum computers. \Cref{sec:algo} details the proposed quantum simulation algorithm, including initial state preparation, circuit construction, and the post-processing steps to obtain the EELS spectrum. \Cref{sec:app} presents an application of the algorithm to compute the EELS spectra of battery materials, including the construction of \ch{Li2MnO3} cluster models, validation with classical simulations, and quantum resource estimates for early fault-tolerant quantum computers. The paper concludes with a summary of findings and an outlook for future applications.

\section{Technical background}\label{sec:background}

The dynamic structure factor is a fundamental quantity that provides a unified description of a wide range of scattering spectroscopies. In this section, we review its definition and properties, focusing on its applications to electron energy loss spectroscopy. 

\emph{Double differential cross section} --- A wide range of spectroscopic experiments can be interpreted in terms of the double differential scattering cross-section $\frac{\partial^2 \sigma}{\partial E \partial \Omega}$, which measures the probability of a probing particle or photon to be scattered into a solid angle $d\Omega$ within the energy interval $[E, E + d E]$, where $dE$ is the width of the interval~\cite{sturm_dynamic_1993}. If collisions occur inelastically, the flux of outgoing particles is proportional to the sum of transition probabilities between the ground and excited states of the target, satisfying energy and momentum conservation. Within the first Born approximation~\cite{sakurai_modern_2020} (valid for high incident energies), the double differential scattering cross-section can be expressed in terms of matrix elements of the interaction Hamiltonian $H_{int}(\mathbf{r},\Lambda)$ coupling the spatial coordinate of the projectile considered a spinless particle in $\mathbb{R}^3$, to the internal degrees of freedom of the target $\Lambda$, where the initial/final states of the former are represented by the plane waves with the wave vectors $\mathbf{k}_{I/F}$, $\langle \mathbf{r}|\mathbf{k}_{I/F}\rangle = (2\pi)^{-3/2} e^{i\mathbf{k}_{I/F}\cdot\mathbf{r}}$. Unless otherwise noted, bold variable notation denotes a vector throughout this manuscript, and atomic units are assumed where $\hbar = e = m_e = 4\pi\epsilon_0 = 1$. Under these assumptions, the double differential cross section $\frac{\partial^2 \sigma}{\partial E \partial \Omega}$ is proportional to the {\it dynamic structure factor}
\begin{align}
    S(\mathbf{q}, \omega) \equiv \sum_{F} |\langle \Phi_I | H_{int}(\mathbf{q}, \Lambda) | \Phi_F\rangle|^2 \delta\left(E_{F}-E_{I}- \omega\right),\label{eq:s_gen} 
\end{align} 
where $\mathbf{q} = \mathbf{k}_F - \mathbf{k}_I$ is the transferred momentum, $\omega$ is the inelastic energy loss, $|\Phi_{I/F}\rangle$ are the internal states of the target with energies $E_{I/F}$, $\delta(\cdot)$ denotes the Dirac delta function, and 
\begin{align}
     H_{int}(\mathbf{q}, \Lambda) = \int d\mathbf{r} \, e^{-i\mathbf{q}\cdot\mathbf{r}} H_{int}(\mathbf{r},\Lambda),
\end{align}
is the Fourier component of $H_{int}(\mathbf{r},\Lambda)$. In this target-centric approach, the physics of the scattering process is fully described by the interaction Hamiltonian. As shown below, depending on the nature of the probe and the details of the energy transfer mechanism, $H_{int}$ can be further simplified. 

\emph{Electron Energy Loss Spectroscopy} --- The formalism outlined above is general and applies to various linear spectroscopies, such as inelastic X-ray scattering, inelastic neutron scattering, and related techniques, as well as electron energy loss spectroscopy~\cite{sturm_dynamic_1993}. Here we consider the latter method in detail. 

The inelastic scattering of free electrons in EELS experiments is governed by their Coulomb interaction with the total charge density of the target system,  $\rho_{tot}(\mathbf{x},\Lambda)$, including both electronic $\{\mathbf{r}_i\}_{1}^{N_e}$ and nuclear degrees of freedom $\{\mathbf{R}_i\}_{1}^{N_{nuc}}$, i.e., $\Lambda = \{\mathbf{r}_i\}_{1}^{N_e} \cup \{\mathbf{R}_i\}_{1}^{N_{nuc}}$ and
\begin{align}
    \rho_{tot}(\mathbf{x},\Lambda) = \sum_{i = 1}^{N_e}\delta(\mathbf{x} - \mathbf{r}_i) + \sum_{j = 1}^{N_{nuc}} Z_j \delta(\mathbf{x} - \mathbf{R}_j), \label{eq:rho_tot}
\end{align}
where $N_e$ and $N_{nuc}$ are the numbers of electrons and nuclei, respectively, and $Z_j$ is the $j$th nuclear charge. While in the most general case, contributions from both the electrons and nuclei need to be accounted for in the interaction Hamiltonian
\begin{align}
    H_{int}(\mathbf{r}, \Lambda) =  \int d\mathbf{x}\, \frac{\rho_{tot}(\mathbf{x},\Lambda)}{|\mathbf{r} - \mathbf{x}|},\label{eq:h_int}
\end{align}
depending on the spectral region, the above expression can be simplified to reflect the dominant energy loss mechanism. An electron energy loss spectrum is typically characterized by three main regions:
\begin{enumerate}
    \item {\it Zero-Loss Peak (ZLP)}. There is an intense peak at zero energy loss due to elastic electron scattering. The ZLP typically features additional fine structure due to phonon excitations in the sample. The phonon modes can be probed in vibrational EELS, either in transmission or the so-called aloof scattering geometry, where the electron beam is oriented parallel to the sample and interacts with it via a long-range electric field (dipole scattering). Aloof EELS is essential for analyzing beam-sensitive materials~\cite{crozier_vibrational_2017}.
    \item {\it Low-Loss Region} ($\omega < 50 $ eV). The prominent spectral features in this region are interband transitions and plasmon peaks, which correspond to collective oscillations of the valence electron gas. 
    \item {\it Core-Loss Region} ($\omega > 50 $ eV). The EELS signal in this region arises from the excitation of atomic core-level electrons into the unoccupied states above the Fermi level. The onset of an excitation edge corresponds to the binding energy of the core electron. It serves as a definitive fingerprint for elemental identification, since the energy separation between core excitations is large between different elements. 
    The fine structure superimposed on the edges provides further chemical information about the sample. Specifically, modulation of the EELS cross-section within about 50 eV of the edge onset, commonly known as ELNES, is highly sensitive to the local chemical environment of the excited atom, including its oxidation state, site symmetry, spin state, and the nature of its chemical bonds.
\end{enumerate}
Core-loss EELS closely resembles XAS while offering the sub-nanometer spatial resolution required to study the local features of heterogeneous materials such as defects, interfaces, and phase boundaries. In particular, K- and L-edge ELNES analysis is one of the primary methods for extracting chemical state information from EELS in battery research~\cite{kimura_stem-eels_2021,zhang_direct_2020,zhang_enhanced_2020,zhang_kinetic_2020,an_distinguishing_2024,yu_stem-eels_2021,nomura_quantitative_2018,castro_characterization_2018} and heterogeneous catalysis~\cite{chee_operando_2023}. 

Using modern scanning tunneling electron microscopes, ELNES spectra can be acquired for energies corresponding to K-edges of the elements up to Zn, L-edges of fifth row elements, and M-edges of lanthanides and actinides~\cite{egerton_electron_2011}. Most recently, even higher energies were accessed in some experiments~\cite{lazar_enabling_2025}. In each case, the final excited electronic states are subject to decay via radiative and non-radiative channels~\cite{agarwal_x-ray_1991}. Since the lifetime broadening of core-excited states does not allow one to resolve the vibrational structure of the core-loss EELS signal, the contribution of the charge density due to nuclei in Eqs.~\eqref{eq:rho_tot} and \eqref{eq:h_int} can be safely ignored. The interaction Hamiltonian then simplifies to
\begin{align}
    H_{int}(\mathbf{r}) = \sum_{i = 1}^{N_e} \frac{1}{|\mathbf{r} - \mathbf{r}_i|}, 
\end{align}
where the summation runs over the coordinates of the electrons only. Therefore, the dynamic structure factor is expressed as
\begin{align}
    S(\mathbf{q}, \omega) = \sum_{F}|\langle \Psi_I| \sum_i e^{i\mathbf{q}\cdot\mathbf{r}_i} | \Psi_F\rangle|^2 \delta\left(E_{F}-E_{I}-\omega\right)\label{eq:s_elnes}, 
\end{align}
where $|\Psi_{I/F}\rangle$ refer to the initial and final {\it electronic} states of the target system in the Born-Oppenheimer approximation. In terms of $S(\mathbf{q}, \omega)$, the double differential scattering cross-sections in the non-relativistic limit is
\begin{align}
\frac{\partial^2 \sigma}{\partial E \partial \Omega}=\frac{4}{|\mathbf{q}|^4} \frac{|\mathbf{k}_F|}{|\mathbf{k}_I|} S(\mathbf{q}, \omega).
\end{align}

Now, to represent the dynamic structure factor in a correlation functional form, it can be shown that the scattering operator $\sum_i e^{i\mathbf{q}\cdot\mathbf{r}_i}$ is the electron density $\hat{n}(\mathbf{r}) = \sum_i \delta(\mathbf{r} - \mathbf{r}_i)$ in reciprocal space:
\begin{align}
    \hat{n}_{\mathbf{q}} = \int d\mathbf{r}\, \hat{n}(\mathbf{r}) e^{i\mathbf{q}\cdot\mathbf{r}} = \sum_i \int d\mathbf{r}\, \delta(\mathbf{r} - \mathbf{r}_i) e^{i\mathbf{q}\cdot\mathbf{r}} = \sum_i e^{i\mathbf{q}\cdot\mathbf{r}_i}.
\end{align}
Following Van Hove in Ref.~\cite{van_hove_correlations_1954}, Eq.~\eqref{eq:s_elnes} can therefore be interpreted as a Fourier transform of the density-density correlation function:
\begin{align}
    S(\mathbf{q}, \omega)  = \frac{1}{2\pi} \int_{-\infty}^{+\infty} dt ~e^{i\omega t} \langle \Psi_I | \hat{n}_{\mathbf{-q}}(t) \hat{n}_{\mathbf{q}} | \Psi_I \rangle, 
\end{align}
where $\hat{n}_{\mathbf{q}}(t)$ is the reciprocal space density operator in the Heisenberg representation, i.e., $\hat{n}_{\mathbf{q}}(t) = e^{iHt}\hat{n}_{\mathbf{q}}e^{-iHt}$. Throughout the rest of this section, we use $H$ to denote the electronic Hamiltonian of the target system with the ground state eigenvector $|\Psi_0\rangle$ and corresponding energy $E_0$. If the initial state is the ground state $| \Psi_I \rangle = |\Psi_0\rangle$, the above expression simplifies to
\begin{align}
    S(\mathbf{q}, \omega) =  \frac{1}{2\pi}\int_{-\infty}^{+\infty} dt \, e^{i\omega t} \langle \Psi_0 | \hat{n}_{-\mathbf{q}} e^{-i(H - E_0)t} \hat{n}_{\mathbf{q}} | \Psi_0 \rangle\label{eq:td_s_elnes}.
\end{align}
To account for the finite lifetime of the core-excited states, we will multiply the correlation function in the above equation by an exponentially decaying factor $e^{-\eta |t|}$, where $\eta$ is a real-valued scalar, chosen to be inversely proportional to the lifetime and $\eta >0$. This is equivalent to replacing each delta function in Eq.~\eqref{eq:s_elnes} by a Lorentzian of width $\eta$ centered at the corresponding transition energy. With this modification, Eq.~\eqref{eq:s_elnes} becomes amenable to numerical simulation and forms the basis of time-dependent approaches for calculating the DSF. 

In anisotropic materials and molecular systems, the DSF is sensitive to the orientation of the $\mathbf{q}$ vector. Anisotropy of $S(\mathbf{q}, \omega)$ provides additional information regarding the electronic structure of the target system~\cite{nelhiebel_theory_1999} and can be detected experimentally in both low-loss~\cite{botton_new_2005} and core-loss~\cite{fossard_angle-resolved_2017,arenal_high-angular-resolution_2007} spectral regions. It is more common, however, to measure the isotropically averaged double differential cross-section, in which case, to model the EELS signal, one needs to integrate the computed DSF over the solid angle $\Omega$ in momentum space
\begin{equation}
    S_{iso}(\mathbf{q}, \omega) = \frac{1}{4\pi} \int d\Omega \, S(\mathbf{q}, \omega).
\end{equation}
In practice, this can be accomplished via numerical integration, by calculating $S(\mathbf{q}, \omega)$ on a grid in momentum space and applying a suitable quadrature rule~\cite{morris_optados_2014}. Whether performed in the time or frequency domain, such calculations require the matrix elements of the one-particle scattering operator in a specific one-particle basis
\begin{equation}
    n_{ij}^{\mathbf{q}} = \int \phi^*_{i} (\mathbf{r}) e^{i \mathbf{q} \cdot \mathbf{r}}\phi_j (\mathbf{r}) \, d\mathbf{r},
\end{equation}
where $\phi_j (\mathbf{r})$ is the $j$th single particle basis function in real-space defined by the electron position $\mathbf{r}$.  For a given $\mathbf{q}$, the matrix element $n_{ij}^{\mathbf{q}}$ is analytically computable for common choices of $\phi_j (\mathbf{r})$, such as Gaussian and plane wave bases, but the average of $S(\mathbf{q}, \omega)$ over the $\mathbf{q}$ variable requires a numerical treatment~\cite{lehtola2012erkale}. This implies that in the general case, the particle scattering integral and its corresponding matrix elements have to be computed for a grid spanning the different $\mathbf{q}$ vectors, which introduces a significant computational overhead.

However, in typical core-loss EELS experiments implemented in STEM, the scattered electrons are detected at small collection angles behind the sample~\cite{egerton_electron_2011}. In this case, the momentum transfer $\mathbf{q}$ is expected to be close to 0, justifying the use  of the {\it dipole approximation}
\begin{align}
    \hat{n}_{\mathbf{q}} \approx \sum_{j = 1}^{N_e} ( 1 + i\mathbf{q}\cdot \mathbf{r}_j)\label{eq:dip_approx}.
\end{align}
The first term corresponds to a non-resonant process (i.e., elastic scattering) and should be excluded from $S(\mathbf{q},\omega)$, while the second term is proportional to the electric dipole moment operator $\bm{\mu} = \sum_{i=1}^{N_e}\mathbf{r}_i$ along the direction of $\mathbf{q}$, and describes the resonant electronic excitations of the target system. The resonant contribution to $S(\mathbf{q},\omega)$ is therefore
\begin{align}
    &S(\mathbf{q}, \omega) \notag \\ 
    &=\frac{-1}{2\pi}\int_{-\infty}^{+\infty} dt~e^{i\omega t-\eta |t|}\langle \Psi_0 | (\mathbf{q}\cdot\bm{\mu}) e^{-i(H - E_0)t} (\mathbf{q}\cdot\bm{\mu}) | \Psi_0 \rangle\\ 
    &= \frac{-1}{2\pi} \sum_{\alpha, \beta} q_{\alpha} q_{\beta} \int_{-\infty}^{+\infty} dt~e^{i\omega t-\eta |t|}\langle \Psi_0 | \mu_{\alpha} e^{-i(H - E_0)t} \mu_{\beta} | \Psi_0 \rangle
\end{align}
where $\alpha, \beta\in \{x,y,z\}$. For convenience, let us introduce
\begin{equation}\label{eq:intensity}
    I_{\alpha\beta}(\omega) \equiv \frac{1}{2\pi} \int_{-\infty}^{+\infty} dt \, e^{i\omega t} e^{-\eta |t|} \langle \Psi_0 | \mu_{\alpha} e^{-i(H - E_0)t} \mu_{\beta} | \Psi_0 \rangle
\end{equation} as the \emph{intensity function}. To compute the dynamic structure factor in the dipole approximation for any $\mathbf{q}$, it suffices to determine a unique $I_{\alpha\beta}(\omega)$. A simple analytic dependence of $S(\mathbf{q}, \omega)$ on $\mathbf{q}$  enables computation of $S_{iso}(\mathbf{q}, \omega)$, in terms of $I_{\alpha\beta}(\omega)$ without resorting to numerical integration over $\mathbf{q}$:  
\begin{align}
S_{iso}(\mathbf{q}, \omega) \propto \frac{|\mathbf{q}|^2}{3} \sum_{\alpha} I_{\alpha\alpha}(\omega).\label{eq:s_dip_iso}
\end{align}
In cases where  $\mathbf{q} \cdot \mathbf{r} \gtrsim 1$, i.e., the norm of $\mathbf{q}$ is larger than the inverse radius of the core hole $r_c$, $|\mathbf{q}|\gtrsim 1/r_c$~\cite{egerton_electron_2011}, the dipole approximation is insufficient and higher-order terms need to be accounted for beyond Eq.~\eqref{eq:dip_approx}~\cite{loffler_breakdown_2011}. Since $r_c$ tends to decrease with the (effective) nuclear charge, dipole-allowed transitions prevail up to relatively high magnitudes of the momentum transfer. However, when beyond-dipole transition effects are non-negligible, they are more challenging to study experimentally since they require larger collection apertures and are generally less intense in the spectrum~\cite{gloter_probing_2009,vos_understanding_2011}. For this reason, beyond-dipole corrections or an exact treatment of the scattering operator are of limited utility for interpreting ELNES in experimentally relevant regimes and will not be further discussed in this work.

\emph{Classical methods for simulating ELNES} --- The interpretation of experimental ELNES spectra often relies on comparing unknown signals to those of reference materials with well-characterized electronic structure and experimental data. This empirical approach is widely used for quantitative elemental analysis, facilitated by the availability of reference libraries~\cite{eelsdb} and advanced software tools for analyzing complex spectra~\cite{pena_hyperspyhyperspy_2025,verbeeck_joverbeepyeelsmodel_2025}. Model-based quantification of ELNES can also be employed to extract chemical information from the fine structure of absorption edges~\cite{verbeeck_model-based_2006}. However, the high variability of these fine structures poses a significant challenge for interpretation~\cite{colliex_early_2022, radtke_energy_2011},  particularly in heterogeneous materials with defects and structural disorder, such as in lithium-ion battery cathodes and heterogeneous catalysts. These difficulties highlight the need for computational methods to enable accurate spectral fingerprinting. 

Since ELNES probes excitations from deep-lying core levels to unoccupied electronic states, accurate theoretical modeling must account for the localized nature of the core hole, electron–hole interactions, and many-body effects. A range of first-principles methods has been developed to address these challenges. 
Conventional approaches based on DFT, including the use of core-hole supercells or transition-state approximations~\cite{ikeno_basics_2017,tait_simulation_2016,gao_core-level_2009}, are often sufficient for light elements and K-edge excitations, particularly when augmented with corrections such as DFT+U or hybrid functionals to treat electron localization. A common workhorse method is TDDFT~\cite{runge1984density}, which solves for approximate particle-hole transition densities in either frequency space~\cite{Besley2010,lopata2011modeling}, or sometimes in the time domain, known as real-time-TDDFT~\cite{bunuau2012time, lopata2011modeling, li2020real}.

For a more accurate treatment of dynamical correlation in core excited states, many-body perturbation theory methods such as the Bethe–Salpeter equation ~\cite{Onida2002} built atop GW quasiparticle corrections~\cite{Hedin1965,Reining2018} are commonly employed~\cite{vinson_bethe-salpeter_2011,ikeno_basics_2017,yao_all-electron_2022}. These methods capture electron-hole interactions and energy-level renormalization, which is critical for reproducing fine spectral features, but they struggle to describe multiplet effects ubiquitous in ELNES, giving rise to missed features or peaks in the simulated spectrum.

Specifically, core excitations in transition metal materials require multi-reference wavefunction-based approaches to incorporate electron–electron interactions and spin–orbit coupling within localized manifolds. Their high computational cost and steep scaling with system size motivate the introduction of the {\it active space} -- a subset of one-particle orbitals relevant to describing the electronic structure of the target states. In most cost-efficient methods, the active space is further subdivided into occupation-restricted sectors, giving rise to so-called RAS approximations. RAS is a natural fit for core-excited states due to their localized nature~\cite{bokarev_theoretical_2020,kasper_ab_2020}. The key aspects of their electronic structure can be understood in terms of atomic-like orbitals hybridized with, or perturbed by, the ligand environment, informing the choice of the orbitals for RAS. Nevertheless, highly accurate calculations require large active spaces beyond the reach of the most advanced methods, such as RASSCF and RASPT2~\cite{fomichev_fast_2025,pinjari_cost_2016,pinjari_restricted_2014}. 

Failure to incorporate necessary orbitals in the active space may result in shifts of transition energies, absence of certain spectral features, and significant discrepancies of relative peak intensities compared to the experiment~\cite{jenkins2021two}. The most advanced RASPT2 simulations of the core-level spectra of transition metal-containing compounds to date were limited to active spaces of thirteen orbitals~\cite{pinjari_cost_2016}. Due to the high complexity of RASSCF/RASPT2 methods and their prohibitive computational cost, simpler model-based alternatives are often adopted in practical applications. These methods encompass charge-transfer multiplet theory~\cite{ikeno_multiplet_2009,haverkort_multiplet_2012,ikeno_basics_2017}, ligand-field DFT~\cite{atanasov_dft_2004}, and further variants~\cite{ramanantoanina_development_2015,de_groot_2p_2021}. The drawbacks of these approaches are primarily due to uncontrolled approximations in the electronic structure, resulting in system-dependent performance. This motivates the need for rigorous electronic structure methods with controllable error, enabling spectral simulations in classically intractable active spaces. Building upon the recent advancements of quantum simulation for XAS~\cite{fomichev_simulating_2024,fomichev_fast_2025}, we propose a new algorithm for ELNES capable of addressing the shortcomings mentioned above. The algorithm is introduced and described in the following section. 

\section{Quantum Algorithm}\label{sec:algo}
In this section, we present a novel quantum algorithm that enables the direct computation of the dynamic structure factor for momentum-resolved spectroscopies, forming the core technical advance of our simulation pipeline (Fig.~\ref{fig:hero}). We start by explaining the main components of computing the intensity in the dipole approximation (related to the dynamic structure factor as defined in the previous section) with respect to initial state preparation and a time-domain approach, which is a generalization of the time-domain algorithm for computing absorption spectra detailed in Ref.~\cite{fomichev_simulating_2024}.

As defined in the previous section, our goal is to compute unique intensity functions, $I_{\alpha\beta}(\omega)$. Since $I_{\alpha\beta}(\omega)$ is the Fourier transform of the time-dependent Green's function $\widetilde{G}_{\alpha\beta}(t)$
\begin{align}
\widetilde{G}_{\alpha\beta}(t) = \langle \Psi_0 | \mu_{\alpha} e^{-i H t} \mu_{\beta} | \Psi_0 \rangle, \label{eq:gf_general}
\end{align}
we choose this as the key expectation value to be computed on the quantum computer. Additionally, $\widetilde{G}_{\alpha\beta}(t)$ needs to be evaluated for each unique pair of Cartesian coordinates $\alpha, \beta$. This amounts to three unique diagonal terms where $\alpha = \beta$, and three unique off-diagonal terms, $\alpha \neq \beta$. As explained below, computing the off-diagonal terms is the most expensive step, because two different dipole operators are needed in the expectation value. 

At a high level, the proposed quantum algorithm for $S(\mathbf{q}, \omega)$ contains the following steps.

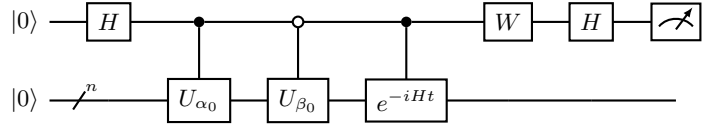
\begin{figure*}
\begin{center}
\begin{quantikz}
\lstick{$|0\rangle$} & \gate{H} & \ctrl{1} & \octrl{1} & \ctrl{1} & \gate{W} & \gate{H} & \meter{}   \\
\lstick{$|0\rangle$} & \qwbundle{n} & \gate{U_{\alpha_0}} & \gate{U_{\beta_0}} & \gate{e^{-i H t}}  &  \qw & \qw & \qw
\end{quantikz}
\end{center}
    \caption{The Hadamard test circuit for evaluating the time-domain Green's function, $\widetilde{G}_{\alpha\beta}(t)$, for a given time $t$ on an  $n$-qubit system register. The one-qubit gate $W = S^{\dagger}\,\,\,\text{or}\,\,\,I$ depending on whether the imaginary or real component of $\widetilde{G}_{\alpha\beta}(t)$ is computed, respectively.}
    \label{fig:had-test-lcu}
\end{figure*}

\begin{enumerate}
\item \textbf{Classical Inputs:} Classically compute the molecular Hamiltonian $H$, the Cartesian components of the dipole operator $\bm{\mu}$, and a classical representation of the initial state $\ket{\Psi_0}$. In the case of electronic structure, this is a linear combination of Slater determinants.  
\item \textbf{Initial State Synthesis:} For each required Cartesian component $\alpha$,  classically apply $\mu_{\alpha}$ to $\ket{\Psi_0}$ to obtain the dipole-rotated state vector $\mu_{\alpha}\ket{\Psi_0}$. Then, from this classical representation, synthesize a unitary $U_{\alpha_{0}}$ that prepares its normalized form, shown in Eq.~\eqref{eq:state_prep_dipole}.
\item \textbf{Time Evolution Synthesis:} Synthesize a unitary that approximates time evolution under the Hamiltonian, $e^{-iHt}$.
\item \textbf{Time-domain Green's Function Evaluation:} For each unique pair of Cartesian directions $(\alpha, \beta)$ and time step $t$, execute the Hadamard test circuit (Fig.~\ref{fig:had-test-lcu}) on a quantum computer. This circuit uses the appropriate controlled state-preparation unitaries ($U_{\alpha_{0}}, U_{\beta_{0}}$) and the time-evolution unitary to measure the real and imaginary parts of the time-domain Green's function $\widetilde{G}_{\alpha\beta}(t)$. (Details of the derivation of this algorithm when $\alpha = \beta$ can be found in Ref.~\cite{fomichev_simulating_2024})
\item \textbf{Spectrum Construction:} Collect the time-domain data and perform a classical discrete Fourier transform on $\widetilde{G}_{\alpha\beta}(t)$ to obtain the intensity functions $I_{\alpha\beta}(\omega)$. Finally, for a chosen set of momentum transfer vectors $\mathbf{q}$, classically combine the computed intensity functions to construct the final dynamic structure factor $S(\mathbf{q}, \omega)$.
\end{enumerate}

We provide a detailed explanation of the algorithm below.

\emph{Initial state preparation} --- As stated above, the quantum algorithm requires the Hamiltonian $H$, Cartesian dipole operator terms, $\bm{\mu}$, and the initial state of the system $\ket{\Psi_0}$. The initial state $\ket{\Psi_0}$ is often assumed to be the ground state (vibrational, electronic, etc.) of some molecular cluster or material system. For example, in many cases this can be a high-quality classical approximation of the electronic ground state, computed using the density matrix renormalization group (DMRG) method~\cite{wouters_density_2014,schollwock_density-matrix_2011} or selected configuration interaction ~\cite{sharma_semistochastic_2017,tubman_deterministic_2016} to find a state in terms of a linear combination of Slater determinants. Then, upon computing a classical description of the state vector, we can create the unitary operator to prepare the state via standard synthesis methods~\cite{fomichev_initial_2024}, or even matrix product state (MPS) based methods for creating compact state preparation circuits~\cite{rudolph_decomposition_2022,ran_encoding_2020,schoen_sequential_2005}. 

Given that the classical representation of the ground state is straightforward to prepare, we also need to compute the action of the dipole operator on this initial state. Naively, one could block encode both dipole terms and apply them sequentially to the initial state before and after the time propagator on the quantum computer directly. However, since we have a quantum state preparation routine for the classical representation of the initial state, the normalized dipole rotated state, $\mu_{\alpha}|\Psi_0\rangle / || \mu_{\alpha}|\Psi_0\rangle ||$ is easy to prepare too, since the classical application of the one-particle dipole operator increases the dimension of $|\Psi_0\rangle$ by a factor of $\mathcal{O}(N_e(N-N_e))$, where $N_e$ is the number of particles, and $N$ is the size of one-particle basis set. We can therefore synthesize a state preparation unitary from the classical representation of this normalized state using the definition
\begin{equation}\label{eq:state_prep_dipole}
    U_{\alpha_0} |0\rangle = \frac{\mu_{\alpha}|\Psi_0\rangle}{|| \mu_{\alpha}|\Psi_0\rangle ||} .
\end{equation}
For brevity, we assume the error in the state preparation unitary is small, and also additive in the final result, and do not explicitly track it in this work.

\emph{Full circuit and postprocessing} --- With the state preparation unitaries, $U_{\alpha_0}$, we can compute the desired expectation value, $\langle 0 |U^{\dagger}_{\beta_0} e^{-i H t} U_{\alpha_0} | 0 \rangle$, assuming we have access to a circuit approximating the time evolution operator. This can be accomplished with the circuit shown in Fig.~\ref{fig:had-test-lcu},
which is a modified Hadamard test using a linear combination of unitaries (LCU) of the two-state preparation circuits with equal weights to capture the off-diagonal components. This approach is agnostic to a particular implementation of the time evolution operator. We typically choose to implement $e^{-iHt}$ via Trotter product formulas due to their reduced qubit cost compared to qubitization-based methods~\cite{low2019hamiltonian}, as detailed in the following section. The choice of a particular circuit for implementing $e^{-iHt}$ will determine the contribution of each Hadamard test to the total runtime. While for a fixed time $t$ the number of ancilla qubit measurements required to estimate the expectation value with precision $\epsilon$ scales as $\mathcal{O}(1/\epsilon^2)$, we have to consider the damping term $e^{-\eta t}$ (Eq.~\eqref{eq:intensity}), and choose the number of samples per $t$ to recover the full Fourier transformed signal $I_{\alpha\beta}(\omega)$ with a given error.

Next, to bound the number of measurements required as well as the maximum $t$ needed in the time evolution circuit, we derive the working expressions for evaluating spectral functions adapted to the case of EELS. Starting from the linear absorption spectroscopy algorithm derivation from Ref.~\cite{fomichev_simulating_2024}, the spectral measure containing the information about excitation energies of a system and its corresponding transition probabilities for each excitation is defined as
\begin{align}
    p_{\alpha\alpha}(x) = \sum_{k=1}^K |\langle \Psi_k | \mu_{\alpha} |\Psi_0\rangle|^2 \delta(x - \tau E_k), ~\text{for}~ x \in [-\pi, \pi), \label{eq:og_spec_measure}
\end{align}
where $K$ is the number of the Hamiltonian eigenstates, $\tau$ is chosen such that $\tau E_k \in [-\pi/2, \pi/2]$, and the function $p_{\alpha \alpha}(x)$ is assumed to be $2\pi$-periodic. This is equivalent to Eq. (7) of Ref.~\cite{fomichev_simulating_2024}. Since this spectral measure is only valid for on-diagonal components, we must generalize Eq.~\eqref{eq:og_spec_measure} to allow for the off-diagonal components required for EELS and the DSF more generally. We define the generalized spectral measure as 
\begin{align}
    p_{\alpha\beta}(x) &= \sum_{k=1}^K \langle \Psi_0 | \mu_{\alpha} |\Psi_k\rangle\langle \Psi_k | \mu_{\beta} |\Psi_0\rangle \delta(x - \tau E_k), \label{eq:spec_meas}
\end{align}
where $p_{\alpha\beta}(x)$ is again assumed to be $2\pi$-periodic. For notational simplicity, we now define $p_{\alpha,k} \equiv \langle \Psi_k | \mu_{\alpha} |\Psi_0\rangle$
to use for the remainder of the paper. 

Now, with this generalized version of the spectral measure, we can derive the intensity functions to compute the on and off-diagonal components of the dynamic structure factor via the modified Hadamard tests, and subsequently the final EELS spectra.

The convolution of $p_{\alpha\beta}(x)$ with a (periodically continued) Lorentzian kernel
\begin{align}
    \mathcal{L}_{\eta}(x) = \frac{1}{\pi} \sum_{n=-\infty}^{\infty} \frac{\eta \tau}{(x - 2\pi n)^2 + (\eta \tau)^2} 
\end{align}
results in
\begin{align}
    C^{\alpha\beta}_{\eta}(x) &= (p_{\alpha\beta}*\mathcal{L}_{\eta})(x)
    = \int_{-\pi}^{\pi} dy\,\, p_{\alpha\beta}(y)\mathcal{L}_{\eta}(x - y)\\   
    &= \frac{1}{\pi} \sum_{k = 1}^{K} \sum_{n=-\infty}^{\infty} \frac{\eta \tau p_{\alpha,k}^*p_{\beta,k}}{(x - \tau E_k - 2\pi n)^2 + (\eta \tau)^2}.
\end{align}
If $(\eta\tau) \ll 1$, the effect of periodic images can be ignored:
\begin{align}
   C^{\alpha\beta}_{\eta}(x) =  \frac{1}{\pi} \sum_{k = 1}^{K}\frac{\eta \tau p_{\alpha,k}^*p_{\beta,k}}{(x - \tau E_k)^2 + (\eta \tau)^2}.
\end{align}
Equivalently, the above expression is
\begin{align}
    C^{\alpha\beta}_{\eta}(x) = \frac{1}{2\pi\tau} \int_{-\infty}^{+\infty} dt \,\, e^{i t\frac{x}{\tau}} e^{-\eta |t|} \widetilde{G}_{\alpha\beta}(t)\label{eq:gf_ft}.
\end{align}
Therefore, in the limit where $\eta \to 0^+$, we recover the exact excitation poles $I_{\alpha\beta}(\omega) = \tau C^{\alpha\beta}_{\eta}(\omega\tau)$.
Leveraging the $2\pi$-periodicity of $p_{\alpha\beta}(x)$ and $\mathcal{L}_{\eta}(x)$, the convolution can be expressed in terms of their Fourier components, providing an exact discretization of the integral in Eq.~\eqref{eq:gf_ft}. Since
\begin{align}
p_{\alpha\beta}(x) = \frac{1}{2\pi} \sum_{j=-\infty}^{+\infty} \widetilde{G}_{\alpha\beta}(j\tau) e^{ijx}
\end{align}
and
\begin{align}
 C^{\alpha\beta}_{\eta}(x) = \frac{1}{2\pi} \sum_{j=-\infty}^{+\infty} e^{-|j|\eta\tau} e^{ijx},   
\end{align}
the convolution theorem yields the following equation for $I_{\alpha\beta}(\omega)$
\begin{align}
    I_{\alpha\beta}(\omega) = \frac{\tau}{2\pi} \sum_{n=-\infty}^{\infty} \widetilde{G}_{\alpha\beta}(n\tau) e^{-|n|\eta\tau} e^{in\tau\omega}\label{eq:spec_func}.
\end{align}
In practice, the sum in the above equation is truncated at $n_{max} =\mathcal{O}( \frac{1}{\eta\tau}\log\frac{1}{\epsilon_{trunc}})$, where $\epsilon_{trunc}$ is the truncation error~\cite{fomichev_fast_2025}. To calculate $S(\mathbf{q}, \omega)$ using this approach, it is convenient to separate its ``diagonal", $S_d(\mathbf{q}, \omega)$, and ``off-diagonal" components, $S_{od}(\mathbf{q}, \omega)$, defined as 
\begin{align}
S_d(\mathbf{q}, \omega) &= \sum_{\alpha}  q_{\alpha}^2 I_{\alpha\alpha}(\omega) \label{eq:S_d},\\
S_{od}(\mathbf{q}, \omega) &= \sum_{\alpha<\beta}  q_{\alpha}q_{\beta} \cdot (I_{\alpha\beta}(\omega) + I_{\beta\alpha}(\omega)).\label{eq:S_od}
\end{align}

The diagonal component $S_d(\mathbf{q}, \omega)$ can be evaluated solely with $\widetilde{G}_{\alpha\alpha}(n\tau)$. To compute the off-diagonal component $S_{od}(\mathbf{q}, \omega)$, it is beneficial to use the symmetries of $\widetilde{G}_{\alpha\beta}(n\tau)$, where ($\alpha\ne\beta$), to reduce the number of queries to a quantum computer. Taking into account that the molecular Hamiltonian is real (which is the case for non-relativistic isolated systems and real-valued one-particle basis sets), we observe that
\begin{align}
    \text{Re}[\widetilde{G}_{\alpha\beta}(n\tau)] &\equiv X^{\alpha\beta}_n =  \langle \Psi_0 | \mu_{\alpha} \cos(Hn\tau) \mu_{\beta} | \Psi_0 \rangle, \\
    \text{Im}[\widetilde{G}_{\alpha\beta}(n\tau)] &\equiv Y^{\alpha\beta}_n = \langle \Psi_0 | \mu_{\alpha} \sin(Hn\tau) \mu_{\beta} | \Psi_0 \rangle,
\end{align}
where we introduced a short-hand notation for the real and imaginary parts of the Green's function, $\widetilde{G}_{\alpha\beta}(n\tau) = X^{\alpha\beta}_n + iY^{\alpha\beta}_n$.
Therefore, $\text{Re}[\widetilde{G}_{\alpha\beta}(n\tau)] = \text{Re}[\widetilde{G}_{\alpha\beta}(-n\tau)]$ and $\text{Im}[\widetilde{G}_{\alpha\beta}(n\tau)] = -\text{Im}[\widetilde{G}_{\alpha\beta}(-n\tau)]$. These relations allow us to simplify the sum $I_{\alpha\beta}(\omega) + I_{\beta\alpha}(\omega)$ and express both $S_d(\mathbf{q}, \omega)$ and $S_{od}(\mathbf{q}, \omega)$ in terms of $X^{\alpha\beta}_n$ and $Y^{\alpha\beta}_n $, which are estimated from the measurement outcomes in the Hadamard test, where the $X$ variable requires $W = I$
and the $Y$ variable requires $W = S^{\dagger}$ in the circuit in Fig.~\ref{fig:had-test-lcu}. The combined total expression for the dynamic structure factor, $S(\mathbf{q}, \omega)$, is then
\begin{widetext}
\begin{align}
    S(\mathbf{q}, \omega) = \sum_{\alpha \ge \beta} q_{\alpha}q_{\beta}\cdot\bigg\{\frac{\tau\cdot(2-\delta_{\alpha\beta})}{2\pi}\cdot \bigg(\langle\Psi_0|\mu_{\alpha}\mu_{\beta}|\Psi_0\rangle + 2\sum_{n=1}^{\infty} [X^{\alpha\beta}_n\cos(n\tau\omega) - Y^{\alpha\beta}_n\sin(n\tau\omega)]e^{-n\eta\tau} \bigg)\bigg\}.\label{eq:s_combined}
\end{align}
\end{widetext}

Here we have shown that the dynamic structure factor can be expressed in terms of real and imaginary parts of $\widetilde{G}_{\alpha \beta}(t)$ for $t > 0$ and $\beta \ge \alpha$. This amounts to evaluating only six intensity functions out of the nine possible, reducing the number of calls to the quantum subroutines for computing the time-dependent Green's functions. Now, splitting the DSF into individual contributions corresponding to unique combinations of $\alpha$ and $\beta$ is computationally efficient when $S(\mathbf{q}, \omega)$ needs to be obtained for several values of $\mathbf{q}$. Alternatively, if only one $\mathbf{q}$ is of interest, $S(\mathbf{q}, \omega)$ can be simulated directly via the spectroscopy algorithm of Ref.~\cite{fomichev_simulating_2024} by preparing and propagating the $\mathbf{q}\cdot\bm{\mu}|\Psi_0\rangle$ state using an LCU approach. 

To complete our analysis, we address the measurement allocation for estimating $S(\mathbf{q}, \omega)$ from the outcomes of the Hadamard tests with the maximum statistical error of $\epsilon_{meas}$.

For the isotropic case, the DSF depends on the norm of $\mathbf{q}$ multiplicatively (Eq.~\eqref{eq:s_dip_iso}), making the EELS cross-section identical to that of XAS up to constant factors. The number of samples needed to measure each $I_{\alpha\alpha}(\omega)$ can therefore be determined similarly, as described in Ref.~\cite{fomichev_fast_2025}. The same is true when $S(\mathbf{q}, \omega)$ is evaluated for a single value of $\mathbf{q}$. 

For the more complicated case of multiple $\mathbf{q}$ vectors, it is more appropriate to estimate all diagonal $I_{\alpha\alpha}(\omega)$ and off-diagonal symmetric combinations of $I_{\alpha\beta}(\omega)$ separately, and account for the $\mathbf{q}$ dependence when combining them into $S_d(\mathbf{q}, \omega)$ and $S_{od}(\mathbf{q}, \omega)$. This strategy ensures that the runtime of the algorithm is independent of the number of $\mathbf{q}$ points. Following the derivation in Ref.~\cite{fomichev_fast_2025}, to measure a single term in the curly braces in Eq.~\eqref{eq:s_combined} with a maximum error of $\delta$, one requires the number of samples $\mathcal{N}_{\alpha\beta}$
\begin{align}
     \mathcal{N}_{\alpha\beta} =  \frac{\tau^2(2-\delta_{\alpha\beta})^2\langle\Psi_0|\mu_\alpha\mu_{\beta}|\Psi_0\rangle^2}{4\pi^2\delta^2}\bigg(\sum_{n'=1}^{n_{max}}e^{-n'\eta\tau}\bigg)^2, \label{eq:sampling}
\end{align}
provided each $\widetilde{G}_{\alpha \beta}(n\tau)$ was estimated using $\mathcal{N}_n$ measurements
\begin{align}
    \mathcal{N}_n = \frac{\mathcal{N}_{\alpha\beta}\cdot e^{-n\eta\tau}}{\sum_{n'=1}^{n_{max}}e^{-n'\eta\tau}}\label{eq:sampling_distr}.
\end{align}
The above sampling distribution is optimal in the sense that it delivers the minimum sampling error $\delta$ subject to the constraint $\sum_{n = 1}^{n_{max}} \mathcal{N}_n = \mathcal{N}_{\alpha\beta}$. Since in the final expression for $S(\mathbf{q}, \omega)$, each contribution is weighted by the product $q_{\alpha}q_{\beta}$, the corresponding variance is proportional to $(q_{\alpha}q_{\beta})^2/\mathcal{N}_{\alpha\beta}$. If $\mathcal{N}_{\alpha\beta}$ are varied to minimize the total variance of $S(\mathbf{q}, \omega)$ for an arbitrary $\mathbf{q}$ in the target set, given a measurement budget $\sum_{\alpha\ge\beta}\mathcal{N}_{\alpha\beta} = \mathcal{N}$, it follows that $\mathcal{N} = \mathcal{O}(|\mathbf{q}|^4/(\eta \epsilon_{meas})^2)$. Taking the maximum over $\mathbf{q}$ then guarantees that $S(\mathbf{q}, \omega)$ will be estimated with an error of less than  $\epsilon_{meas}$ for all of the target values of $\mathbf{q}$. Since the double differential scattering cross-section is proportional to $S(\mathbf{q},\omega)/|\mathbf{q}|^4$ the number of measurements required to achieve the error $\epsilon_{meas}^{\prime}$ in $\frac{\partial^2 \sigma}{\partial E \partial \Omega}$ is $\mathcal{O}(1/(|\mathbf{q}|^2\eta \epsilon_{meas}^{\prime})^2)$.

\section{Application}
\label{sec:app}
The study of \ch{Li2MnO3} has played a central role in advancing the understanding of oxygen redox processes in lithium-rich layered oxides, which is a class of cathode materials widely researched for their promise to increase the total energy capacity of lithium-ion batteries. As one of the prototypical components in composite cathodes, \ch{Li2MnO3} possesses a monoclinic C2/m structure (Materials Project~\cite{jain_commentary_2013} ID mp-18988) and has unique cation arrangement, with excess lithium occupying transition‑metal layers, creating Li-O-Li bonding environments that enable reversible participation of lattice oxygen in the charge‑compensation process beyond the conventional transition‑metal redox~\cite{zhang_pushing_2022}. This makes \ch{Li2MnO3} an ideal model system to investigate the fundamental mechanisms of oxygen redox, including oxygen hole formation, O-O dimerization, and potential oxygen loss, as well as the associated structural and voltage hysteresis phenomena ~\cite{grimaud_anionic_2016}. Understanding oxygen redox behavior in \ch{Li2MnO3} provides key insights for the rational design of next‑generation cathodes that leverage oxygen redox to achieve capacities exceeding the limits imposed by transition‑metal redox alone, while mitigating voltage fade and interfacial instability.
\begin{figure*}
    \centering
    \includegraphics[width=\textwidth]{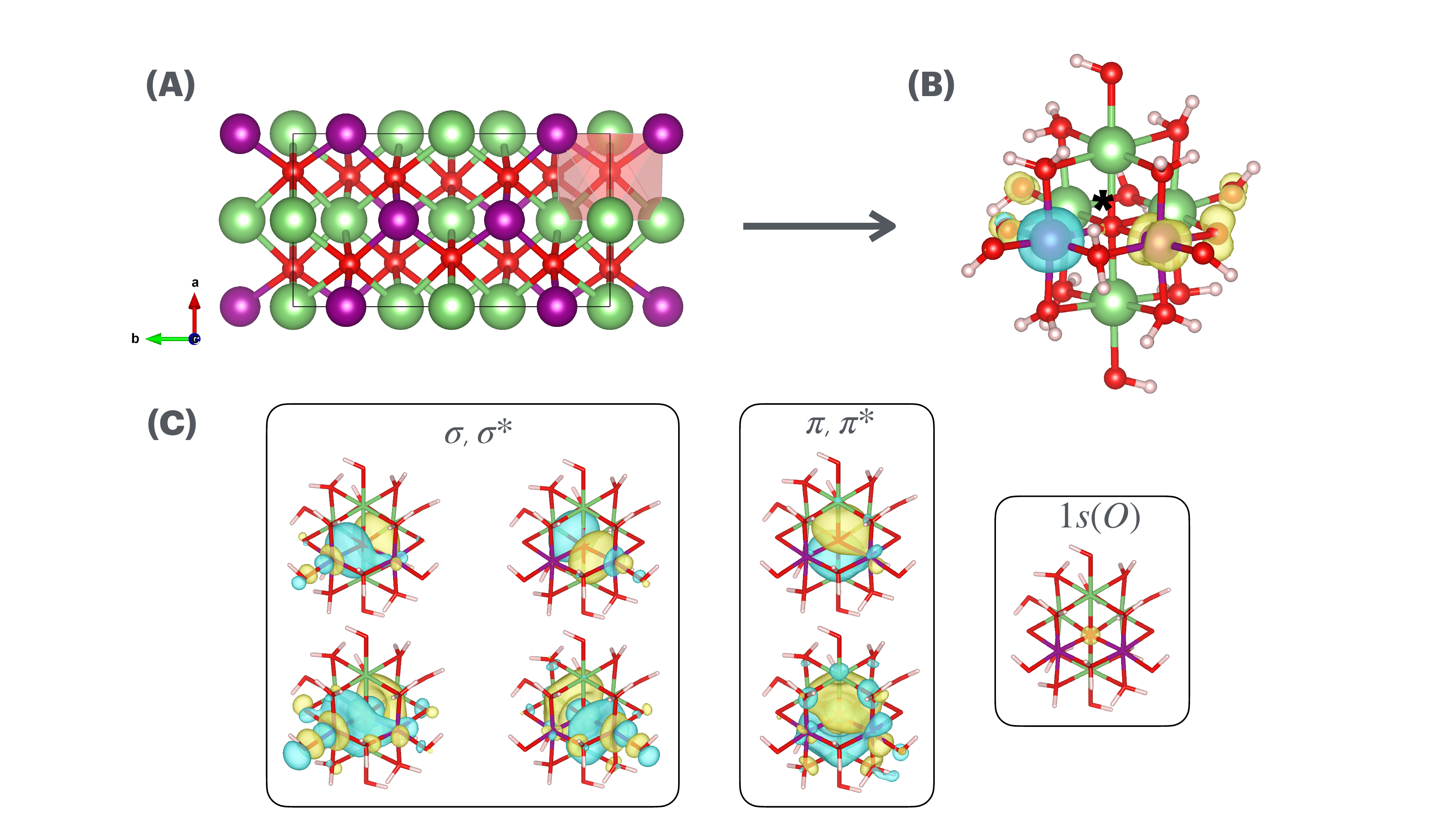}
    \caption{(A) Unit cell of the pristine \ch{Li2MnO3} crystal (C2/m space group) with an oxygen-containing \ch{Li4Mn2O} unit shown with the red octahedron; (B) The cluster model of \ch{Li4Mn2O} used throughout this section with the empirical formula \ch{Li4Mn2O19H26}. Additionally, the spin density of the $S_z = 0$ ground state is superimposed (computed at the PBE0/ANO-RCC-VDZP level of theory in PySCF~\cite{sun_pyscf_2018}). Positive (negative) values of the spin density are indicated with yellow (blue) isosurfaces. The expected absolute values of the integrated spin-density are $3/2$, in line with antiferromagnetic coupling of high-spin Mn ($3d^3$) centers; (C) The isosurfaces of the orbitals spanning the CAS(8,7) active space, centered on the oxygen atom marked with a black asterisk in B. The $\sigma$-type orbitals describe bonding between the central O and Mn atoms, while the $\pi$-type orbitals are localized on O and are weakly hybridized with the Mn $3d$ shell. The $1s \to \pi^*$ and $1s \to \sigma^*$ transitions are the principal contributions to the O K-edge ELNES.} 
    \label{fig:clusters}
\end{figure*}

Experimental advances, particularly in resonant inelastic X-ray scattering, hard and soft XAS, solid-state nuclear magnetic resonance~\cite{serrano-sevillano_dft-assisted_2019}, and high-resolution STEM, have contributed to a molecular-level understanding of the charging mechanisms in \ch{Li2MnO3} and related battery materials. Recent work suggested, however, that defects~\cite{reynaud_imperfect_2023,serrano-sevillano_impact_2021} (such as stacking faults) and structural disorder~\cite{qiu_negative_2025} are highly correlated with their electrochemical performance. For this reason, spatially resolved STEM/EELS techniques can provide crucial details of the redox chemistry, and have been successfully used for battery cathode studies in the past~\cite{yu_stem-eels_2021}.
Here, we demonstrate how quantum simulation can address the need for a robust interpretation tool for ELNES spectra of battery materials by showcasing an application of our algorithm to simulate the oxygen K-edge ELNES spectrum of pristine \ch{Li2MnO3}.

\emph{Cluster models and active space selection} --- Due to the spatially localized nature of core excitations, excited states contributing to oxygen K-edge ELNES can be efficiently simulated using molecular cluster models. In \ch{Li2MnO3} (C2/m), oxygen atoms reside within octahedral  \ch{Li4Mn2O} units. To minimize finite-size effects, a corresponding cluster model was prepared by carving the oxygen-centered coordination polyhedron, along with its immediate neighboring Li and Mn atoms, out of the original crystal structure. The dangling bonds of the outer shell were saturated with hydrogen atoms as in previous XAS simulations of transition-metal oxide materials reported in the literature~\cite{maganas_first_2013,kolczewski_identification_2003,staemmler_cluster_2005}. The resulting cluster with an empirical formula of \ch{Li4Mn2O19H26} is shown in Fig.~\ref{fig:clusters}B., and all classical electronic structure calculations were performed with the PySCF~\cite{sun_pyscf_2018} software. 

In \ch{Li2MnO3}, Mn atoms have a formal oxidation state of $+4$ $(3d^3)$, with three $d$-electrons occupying the $t_{2g}$ orbitals which gives rise to the local spin of $S = 3/2$. The Mn centers within and between the honeycomb layers (along the \emph{a}-axis in Fig.~\ref{fig:clusters}A), are coupled anti-ferromagnetically~\cite{lee_antiferromagnetic_2012}. Based on periodic DFT calculations with GGA+U type exchange-correlation functionals~\cite{xiao_density_2012}, the top of the valence band correlates with the O $2p$ orbitals, whereas the bottom of the conduction band is dominated by the hybridized O $2p$, and Mn $3d$ orbitals. 

The ground state electronic structure of the cluster model constructed in this work is qualitatively similar to the expected formal oxidation and spin states, as well as the orbital character exhibited in previous studies of pristine \ch{Li2MnO3}. In particular, it exhibits antiferromagnetic coupling of Mn spins in its ground state computed with unrestricted PBE0 \cite{adamo1999toward} and the ANO-RCC-VDZP basis set \cite{roos2004main,roos2004relativistic,roos2005new}, using a broken symmetry initial guess for $S_z = 0$. This spin ordering gives rise to a non-zero spin density delocalized over the two Mn atoms in the \ch{Li4Mn2O} unit, as shown in Fig.~\ref{fig:clusters}B. 

Having established the validity of the cluster model, we proceeded to generate a series of effective Hamiltonians by reducing the initial set of molecular orbitals to active subspaces of varying size and complexity. 

In the one-electron picture, the O K-edge ELNES spectrum of \ch{Li2MnO3} involves transitions from the $1s$ core orbital of O to its $2p$ orbitals hybridized with the $3d$ orbitals of Mn. To construct a minimal model describing the relevant excitation mechanism, it is therefore necessary to include the O $1s$ and $2p$ orbitals with a total population of 8 electrons in the active space. This model can be expanded to include Mn $3d$ orbitals to capture the static electronic correlation in the Mn $d$-shell. An even further increase in the active space size can be necessary to account for dynamic correlation~\cite{pinjari_restricted_2014} or non-trivial aspects of transition metal electronic structure, such as the double-shell effect~\cite{pierloot_caspt2_2003}. 

To explore these chemically motivated active space choices, we applied the Atomic Valence Active Space (AVAS) approach of Sayfutyarova et al.~\cite{sayfutyarova_automated_2017} with a tunable threshold. In AVAS, the active space is spanned by the orbitals selected among the eigenvectors of a projector on a user-defined set of reference atomic orbitals in a minimal basis. Once the active orbitals $\{\phi_p\}_1^{N_a}$ are defined, the electronic molecular Hamiltonian is expressed in the second-quantized form
\begin{align}
H = H_0 \, + & \sum_{p, q=1}^{N_a} \sum_{\sigma \in\{\uparrow, \downarrow\}}(p|\kappa| q) a_{p \sigma}^{\dagger} a_{q \sigma}\\
+& \frac{1}{2} \sum_{p, q, r, s=1}^{N_a} \sum_{\sigma, \sigma^{\prime} \in\{\uparrow, \downarrow\}}(p q | r s) a_{p \sigma}^{\dagger} a_{q \sigma} a_{r \sigma^{\prime}}^{\dagger} a_{s \sigma^{\prime}}, \label{eq:mol_ham}
\end{align}
where $a_{p \sigma}^{\dagger}$ ($a_{p \sigma}$) is the creation (annihilation) operator for spatial orbital $p$ and spin $\sigma$, $H_0$ is the energy offset due to the occupied orbitals not included in the active space, and $(p|\kappa| q)$ and $(p q | r s)$ are one- and two-body integrals in chemist's notation respectively, defined as 
\begin{align}
(p q | r s) & \equiv \iint d \mathbf{r}_1 d \mathbf{r}_2 \phi_p\left(\mathbf{r}_1\right) \phi_q\left(\mathbf{r}_1\right) \frac{1}{r_{12}} \phi_r\left(\mathbf{r}_2\right) \phi_s\left(\mathbf{r}_2\right), \\
(p|\kappa| q) & \equiv(p|-\frac{1}{2} \nabla_1^2 + v_1^{HF}-\sum_A \frac{Z_A}{r_{1 A}}| q) \notag\\
& -\frac{1}{2} \sum_r(p r \mid q r),
\end{align}
with $v_1^{HF}$ being the Hartree-Fock potential of inactive electrons~\cite{szabo_modern_1996}, $Z_A$ being nuclear charge of nucleus $A$ and $r_{ij}$ ($r_{1A}$) inter-electron (electron-nucleus) distances, and $\nabla^2$ is the spatial Laplace operator.  

To analyze the resource requirements of our quantum algorithm and validate the choices for optimal simulation settings, we defined an eight-electron/seven-orbital active space for the cluster model of \ch{Li2MnO3}, by selecting the $1s$ and $2p$ orbitals centered on the oxygen atom marked with a black asterisk in Fig.~\ref{fig:clusters}B, and setting the AVAS threshold to 0.09~\cite{sayfutyarova_automated_2017}. We will refer to this active space as CAS(8,7) to indicate that no occupation restrictions were imposed on the constituent orbitals, making it a {\it complete active space}~\cite{roos_complete_1987,bokarev_theoretical_2020} (CAS).

The shapes of the orbitals in the CAS(8,7) active space are shown in Fig.~\ref{fig:clusters}C. They describe $\sigma$ bonds between the central O atom and the two Mn atoms, as well as a lone pair on O, weakly hybridized with the $d$ orbitals on Mn to form $\pi$-type bonding and anti-bonding combinations. This shows that the CAS(8,7) active space captures key qualitative features of the $1s\to2p$ excitation contributing to the O K-edge~\cite{radtke_energy_2011}. To improve the description of the relevant many-body states, CAS(8,7) can be expanded to include $3d$ orbitals of Mn, giving rise to a fourteen-electron/fourteen-orbital active space, CAS(14,14). While the former is spanned by  $\sim 10^3$ Slater determinants within the $S_z=0$ sector,  the latter has the dimension of $\sim 1.2\times10^7$, making it challenging~\cite{delcey_efficient_2019} to perform direct diagonalization of $H$, without imposing additional approximations~\cite{pinjari_cost_2016}. Quantum simulation of the O K-edge ELNES, however, remains viable beyond the classical tractability regime, as described below.

\emph{Parameters for quantum simulation} -- To evaluate the dynamic structure factor and ultimately the ELNES cross section, as prescribed by Eq.~\eqref{eq:s_combined}, one needs to determine (a) the truncation level of the Fourier series, $n_{max}$; (b) the number of samples to evaluate $\widetilde{G}_{\alpha\beta}(n\tau)$ with $n \in [1, n_{max}]$ for the Hadamard tests; and (c) implementation and compilation details of the time-evolution operator $e^{-iHt}$. The interplay of the errors due to the approximations involved in each step underscores the importance of classical simulation in choosing the optimal settings for the quantum simulation.

We choose to implement $e^{-iHt}$ via Suzuki-Trotter product formulas~\cite{suzuki_general_1991,childs_theory_2021} combined with the Compressed Double Factorization (CDF)~\cite{cohn_quantum_2021} and the Block-Invariant Symmetry Shift~\cite{loaiza_reducing_2023,loaiza_block-invariant_2023} as described in Ref.~\cite{fomichev_fast_2025}. The accuracy of the CDF representation of $H$ is controlled by the \emph{factorization rank $L$}, which was set to the number of spatial orbitals. Within this framework, the time evolution subroutine required for calculating $\widetilde{G}_{\alpha\beta}(n\tau)$ was implemented as follows
\begin{align}
    e^{-in\tau H} \approx\bigg[[U_{2}(\tau/k)]^k\bigg]^n,\label{eq:trotter}
\end{align}
where $U_{2}(t)$ is the second-order Suzuki-Trotter product formula~\cite{suzuki_general_1991} for the propagator. In Eq.~\eqref{eq:trotter}, the Hamiltonian was represented in the second quantization (Eq.~\eqref{eq:mol_ham}) using the Jordan-Wigner mapping of spin-orbitals to qubits.
The value of $k$ was chosen independently of the total simulation time~\cite{fomichev_fast_2025} to ensure the accuracy of the spectral features in the calculated dynamic structure factor $S(\mathbf{q}, \omega)$. With this approach, the quantum resources required to compute $\widetilde{G}_{\alpha\beta}(n\tau)$ for $n$ up to $n_{max} = \widetilde{\mathcal{O}}(\frac{1}{\eta\tau})$ in Eq.~\eqref{eq:spec_func} scale asymptotically as $\widetilde{\mathcal{O}}(\frac{1}{\eta^2\tau})$ up to logarithmic factors. Therefore, $\tau$ has to be maximized for resource-efficient simulation. 

To avoid nonphysical aliasing effects in the periodically continued spectral measure $p_{\alpha\beta}(x)$ (Eq.~\eqref{eq:spec_meas}),  $\tau$ can be chosen as $\frac{\pi}{2||H||}$, where $||H||$ is the spectral norm of the Hamiltonian $H$. By imposing \emph{core-valence separation} (CVS)~\cite{norman_simulating_2018} on the Hamiltonian $H$ and the dipole operator $\bm{\mu}$, it is possible to restrict $||H||$ to the energy window containing the target core-excited states~\cite{fomichev_simulating_2024}. The CVS is a standard approximation in classical simulation of core-level spectroscopies, and its effect on the target states is known to be negligible~\cite{herbst_quantifying_2020}. Using CVS is essential for the efficiency of our algorithm, since $\tau$ can be set to $\frac{\pi}{\Delta}$, where $\Delta$ is the energy window that spans the oxygen K-edge spectrum for the cluster model of \ch{Li2MnO3}. This choice is independent of the active space size and applies to oxygen K-edges in other systems due to the locality of core excitations and element specificity. 

In addition to the computational overheads associated with implementing the time evolution operator on a quantum computer, we must also evaluate the sampling cost. To estimate one instance of the time-domain Green's function  $\tilde{G}_{\alpha\beta}(t)$ with precision $\epsilon$, one needs to apply the circuit in Fig.~\ref{fig:had-test-lcu} and measure the ancilla register $\mathcal{O}(1/\epsilon^2)$ times. As shown at the end of the previous section, the required accuracy is fixed by time $t$, the width parameter $\eta$, and the target precision of the spectral function. In practice, one can tolerate a fraction of sampling noise in the latter, as long as its magnitude allows one to unambiguously resolve the spectral features.  

\begin{figure}[!h]
    \centering
    \includegraphics[width=0.4\textwidth]{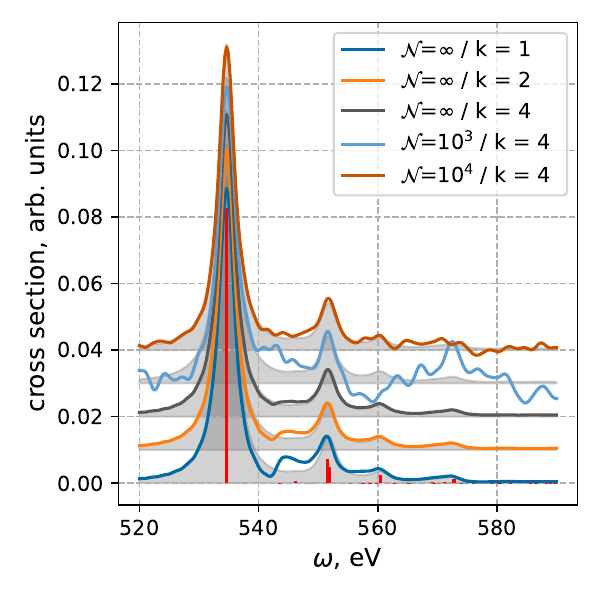}
\caption{The oxygen K-edge ELNES cross section for the cluster model of \ch{Li2MnO3} (Fig.~\ref{fig:clusters}B) in the CAS(8,7) active space for $\mathbf{q} = (1, 1, 1)$. Vertical red lines are placed at the excitation energies corresponding to the eigenvalues of the electronic Hamiltonian. Their heights are proportional to the squares of the projected transition dipole moments, $|\langle\Psi_0 | \mathbf{q}\cdot\mathbf{\mu}|\Psi_i\rangle|^2$. $\mathcal{N}$ is the measurement budget, and $k$ refers to the Trotter step in the approximate time evolution operator from Eq.~\eqref{eq:trotter}.  The grey spectra superimposed on the colorful lines correspond to the exact result for the CAS(8,7) active space. Colorful lines were obtained by simulating the ELNES algorithm in Pennylane~\cite{bergholm_pennylane_2022}. The CDF rank used was $L$ = 7, $\Delta$ = 3.28 Ha, $\eta = 0.06$ Ha, $n_{max} = 87$, and $\tau = 0.096$. Simulation results are offset vertically for clarity.} 
    \label{fig:cl_eels}
\end{figure}

In Fig.~\ref{fig:cl_eels}, we illustrate the interplay between different sources of error in the O K-edge ELNES of our cluster model in the CAS(8,7) active space. As expected, all of the spectra exhibit two intense bands originating from $1s \to \pi^*$ and $1s \to \sigma^*$ transitions. 
The simulated spectral functions based on Eq.~\eqref{eq:s_combined} agree with the reference computed from the energies and the transition dipole moments obtained with CASCI in PySCF~\cite{sun_pyscf_2018}, which, within the active space, is the exact result. Each simulation was performed by preparing $\mathbf{q}\cdot\bm{\mu}|\Psi_0\rangle$, where the ground state $|\Psi_0\rangle$ was represented with the lowest energy CASCI eigenvector, and a CDF factorization rank of $L = 7$.

In the noiseless simulations (i.e., the spectra with the total number of shots $\mathcal{N} = \infty$), the unphysical band around $545$ eV disappears as one systematically decreases the Trotter time step. This confirms that the chosen parameters -- $\tau = \frac{\pi}{\Delta}$, where $\Delta \approx 3.3$ Ha, and $n_{max} \approx 5/(\eta\cdot\tau) = 87$ -- are sufficient to obtain an accurate spectrum, provided the Trotter step is less than or equal $\tau/4$. Furthermore, about $10^4$ samples are needed to capture accurate spectral features in the presence of statistical error. 

\emph{Quantum resource estimates} -- To assess the potential of our ELNES simulation algorithm to deliver quantum advantage, we evaluate logical resource requirements for problem sizes at the fringe, and beyond that of classical tractability. As representative model active space sizes, we chose the number of orbitals varied from 14 to 30. Further, we assumed that the parameters found via classical simulation of the algorithm for the CAS(8,7) active space are close to optimal for larger problem instances, and one needs to calculate $S(\mathbf{q},\omega)$ for a single value of $\mathbf{q}$.   

A series of optimizations was applied when estimating the number of T gates required for the time evolution, as described in Refs.~\cite{fomichev_fast_2025,loaiza_simulating_2025}. Controlled arbitrary angle rotations were replaced with uncontrolled versions at the expense of introducing extra Clifford gates, effectively reducing the number of T gates needed to implement the controlled time evolution by a factor of 2. Following Ref.~\cite{fomichev_fast_2025}, the accuracy of single qubit rotation synthesis $\epsilon_r$ was set to $10^{-3}$, both for state preparation and the controlled time evolution. The number of T gates per single $Z$ rotation was estimated as
\begin{align}
C_{\text {rot }}\left(\epsilon_{r}\right)=0.53 \log _2\left(\epsilon_r^{-1}\right)+4.86,
\end{align}
from Ref.~\cite{kliuchnikov_shorter_2023}.
The time step in the second-order Trotter formula was set to $\tau/4$, which closely corresponds to the perturbation theory estimate $\sqrt{\eta}$~\cite{fomichev_fast_2025}, and agrees with the value found by simulating the algorithm for the CAS(8,7) active space (Fig.~\ref{fig:cl_eels}). To assess the resources required for initial state preparation, we estimated the T gate cost of synthesizing $U_{\alpha_0}$ (Eq.~\eqref{eq:state_prep_dipole}) for the ground state of the cluster model in the larger CAS(14,14) active space. 

We now describe in detail the quantitative resource estimate results. As shown in Fig.~\ref{fig:had-test-lcu}, the state preparation block of the Hadamard test circuits for computing $S_{od}(\mathbf{q},\omega)$ includes two applications of controlled state preparation unitaries, in contrast to $S_{d}(\mathbf{q},\omega)$, where only a single uncontrolled state preparation is required. To analyze the T gate and qubit footprint in each case, we apply the Sum of Slaters (SOS) approach~\cite{fomichev_initial_2024} to synthesize $U_{\alpha_0}$ (Eq.~\eqref{eq:state_prep_dipole}). The ground state of the cluster was classically computed with the spin-adapted DMRG method at a maximum bond dimension 100 using \texttt{block2}~\cite{zhai_block2_2023}. The resulting DMRG ground state was converted to a full state vector by sampling the determinants with amplitudes larger than $10^{-3}$ from the optimized matrix product state. The resulting vector contained 9944 Slater determinants and had the ground state energy shifted by $\sim$ 0.3 eV with respect to the original DMRG solution. Upon action of the dipole operators, the dimension of the state vector increased to about $1.6 \times 10^4$ -- $1.8 \times 10^4$. Using these states as classical inputs for synthesizing $U_{\alpha_0}$, we found the cost of the state preparation block to be $1.1\times10^6$ and $9.7\times10^6$  T gates, for $S_{d}(\mathbf{q},\omega)$ and $S_{od}(\mathbf{q},\omega)$, respectively. Comparing the state preparation costs to those in Tab.~\ref{tab:resources}, we observe that the time evolution dominates the total resource requirements for the quantum simulation algorithm.

As noted above, taking the DMRG solution and projecting it onto the subspace of dominant Slater determinants introduces a bias in the ground state energy. However, this only uniformly shifts absolute transition energies by a value smaller than the lifetime broadening of the O 1s core-excited states ($\eta = 0.06~Ha\approx1.6~eV$), and can be safely neglected. Beyond errors in peak positions themselves, contamination of the approximate ground state with excited state eigenvectors may have more severe consequences, resulting in aliasing artifacts in the computed spectral function. However, the intensity of the nonphysical peaks decreases proportionally to $1 - \langle\Psi_{init}|\Psi_0\rangle$ and is expected to be negligible for the above example, since $\langle\Psi_{init}|\Psi_0\rangle \approx 0.992$. To carry out resource estimation for larger active spaces, we therefore assumed that the ground state can be represented with no more than $10^4$ Slater determinants and can be prepared with the SOS approach. 

The final quantum resources requirements to compute $S(\mathbf{q}, \omega)$ for this oxygen K-edge ELNES application are reported in Tab.~\ref{tab:resources}. These values are reported in terms of the T gate cost of the deepest Hadamard circuit (labeled ``Deepest circuit") and the total number of T gates across all circuits, accounting for the total number of repetitions $\mathcal{N}_n$ (see Eq.~\eqref{eq:sampling_distr}) using $\mathcal{N} = 10^4$ (labeled ``Algorithm"). As expected from the properties of CDF, the total computational cost  (presented in the ``Algorithm" section of Tab.~\ref{tab:resources}) scales approximately as $\mathcal{O}(N_a^3)$, ranging from $1.5 \times 10^{12}$ to $1.5 \times 10^{13}$ for the smallest and largest problem instances, respectively. Compared to the XAS resource requirements reported in Ref.~\cite{fomichev_fast_2025} for $N_a = 18$, we note that our algorithm requires about $7\times$ more T gates due to the need to compute off-diagonal intensity functions as well as a higher sampling cost and larger $n_{max}$ 
for our model of \ch{Li2MnO3}. 

T gate counts serve as a proxy for the algorithm's runtime. To convert them into physical resource estimates, such as runtime, the quantum circuits for computing $\widetilde{G}_{\alpha\beta}(t)$ need to be compiled for a particular quantum processing unit architecture. Here, we employed the active volume compilation approach, introduced in Ref.~\cite{litinski_active_2022}, to estimate runtimes given a single fault-tolerant quantum computer. Active volume compilation affords a significant degree of parallelism when implementing circuit operations, allowing for the reduction of circuit depth. Based on the elementary building blocks of a quantum circuit, one can compute its active volume $V$ as prescribed by Table I of Ref.~\cite{litinski_active_2022} (using the cost of T gates only) and estimate the circuit depth as 
\begin{align}
    \text{Depth}\approx \frac{2V}{n_q},
\end{align}
where $n_q$ is the number of available logical qubits. 

The values of the active volume for our algorithm vary between $4.13\times10^{13}$ and $4.08\times10^{14}$ depending on $N_a$ as shown in Tab.~\ref{tab:resources}. Taking $N_a = 22$ as an example, and assuming a total qubit budget $n_q = 350$, we estimated the corresponding circuit depth to be $9.14\times10^{11}$. On a fault-tolerant quantum computer where non-Clifford gates are the bottleneck, with a 1 MHz clock cycle, this corresponds to the runtime of 10.5 days.

\section{Conclusion}

This work establishes a comprehensive quantum simulation framework capable of computing the dynamic structure factor, a foundational quantity for interpreting a wide range of inelastic scattering experiments with a variety of scattering probes (photons, electron beams, etc.). We achieve this by introducing a novel time-domain quantum algorithm that generalizes previous methods for linear absorption spectroscopy,~\cite{fomichev_fast_2025,fomichev_simulating_2024} to include the off-diagonal terms of the time-domain Green's function, and include the momentum transfer vector of the scattering probe. This capability is essential for modeling orientation-dependent and momentum-resolved spectra. To demonstrate the utility of our approach, we applied it to a significant challenge in materials science: the simulation of core-loss electron energy loss spectroscopy (EELS) for battery cathode materials. This integrated approach significantly broadens the applicability of quantum simulation, providing a path for the quantitative interpretation of spectroscopic data from various types of experiments.

To demonstrate the practical utility of our method, we analyzed and provided quantum resource estimates for simulating the oxygen K-edge ELNES spectrum of a cluster model of lithium manganese oxide \ch{Li2MnO3}; a prototypical cathode material where understanding local electronic structure is crucial for improving battery performance. Another key contribution of our work was in the construction of an appropriate electronic structure model for \ch{Li2MnO3}, which is fundamental to the physical validity of the final simulated spectra. To this end, we developed a cluster model derived directly from the experimental crystal structure. To ensure the physical relevance of our simulations, the derived cluster model was validated against known properties of the material. We confirmed that our model correctly reproduces the essential antiferromagnetic ground-state ordering of the manganese ($3d^3$) centers, as well as the appropriate qualitative character of the valence orbitals surrounding the central oxygen atom that contains the core electron included in the active space model. Appropriately capturing these properties is a prerequisite for generating a physically meaningful electronic structure and, consequently, a reliable EELS spectrum to serve as a basis for interpreting experimental data.

Additionally, through detailed classical simulations and analysis, we established a set of optimized parameters for the quantum algorithm, demonstrating that it can accurately reproduce spectral features with a moderate number of measurements and a feasible Trotter step size for the Hamiltonian simulation subroutine, placing the accurate simulation of these classically challenging systems within the reach of early fault-tolerant quantum computation. 

Future research directions include extending our results to simulating other momentum-resolved spectroscopies on quantum computers, and also the theoretical extension of the DSF formalism to incorporate higher-order multipole contributions beyond the dipole approximation for simulating more exotic scattering experiments.

\section*{Acknowledgments}
We thank Tarik El-Khateeb for assistance with creating Fig. ~\ref{fig:hero}, Pablo Antonio Moreno Casares, Danial Motlagh, and Ignacio Loaiza for early feedback and helpful discussions. This research used resources of the National Energy Research Scientific Computing Center (NERSC), a Department of Energy User Facility using NERSC award DDR QIS-ERCAP 0032729.

\bibliographystyle{unsrt}

\end{document}